\documentclass[useAMS,usenatbib]{mn2e}
\usepackage{epsfig}

\title{The environments of hyperluminous infrared galaxies at $0.44<z<1.55$}
\author[D.~Farrah et al.]
 {D.~Farrah,$^1$\thanks{E-mail: duncan@ipac.caltech.edu}
 J.~Geach,$^{2,3}$
 M.~Fox,$^{2}$
 S.~Serjeant,$^{4}$
 S.~Oliver,$^{5}$
 A.~Verma,$^{6}$ \newauthor
 A.~Kaviani,$^{2}$ and
 M.~Rowan-Robinson$^{2}$\\
 $^{1}$\it SIRTF Science Center, Jet Propulsion Laboratory,
 California Institute of Technology, Pasadena 91125, USA\\
 $^{2}$\it Astrophysics Group, Blackett Laboratory, Imperial College, Prince Consort
 Road, London SW7 2BW, UK\\
 $^{3}$\it Astronomy Unit, Queen Mary College, Mile End Road, London E1 4NS, UK\\
 $^{4}$\it Centre for Astrophysics and Planetary Science, School of Physical Sciences, University of Kent, Canterbury, Kent, CT2 7NR, UK\\
 $^{5}$\it Astronomy Centre, University of Sussex, Falmer, Brighton BN1 9QJ, UK\\
 $^{6}$\it Max-Planck-Institut fur Extraterrestrische Physik, Postfach 1312, 85741 Garching, Germany}

\pagerange{\pageref{firstpage}--\pageref{lastpage}} \pubyear{2003}

\begin{document}

\maketitle

\label{firstpage}
\begin{abstract}
We present deep wide-field $K_{s}$-band observations of six Hyperluminous Infrared Galaxies (HLIRGs) spanning a
redshift range $0.44<z<1.55$. The sample resides in a wide variety of environments, from the field to Abell 2
clusters, with a mean galaxy-HLIRG clustering amplitude of $\langle B_{gh} \rangle=190 \pm 45$Mpc$^{1.77}$. The range
in environments, and the mean clustering level, are both greater than those seen in local IR-luminous galaxies,
from which we infer that the range of galaxy evolution processes driving IR-luminous galaxy evolution at
$z>0.5$ is greater than locally, and includes mergers between gas-rich spiral galaxies in the 
field, but also includes encounters in clusters and hierarchical buildup.  The similarity in the range of 
environments and mean clustering amplitude between our sample and QSOs over a similar redshift range is 
consistent with the interpretation where evolutionary connections between IR-luminous galaxies and QSOs are 
stronger at $z>0.5$ than locally, and that, at these redshifts, the processes that drive QSO evolution are 
similar to those that drive IR-luminous galaxy evolution. From comparison of the HLIRG and QSO host galaxies 
we further postulate that a larger fraction of IR-luminous galaxies pass through an optical QSO stage at 
$z>0.5$ than locally. 
\end{abstract}

\maketitle  

\begin{table*}
\caption{Hyperluminous Infrared Galaxy sample \label{hlirgsample}}
\begin{tabular}{@{}lcccccccccc@{}}
\hline
Name             & $z$  & RA (J2000) & Dec        & Spectrum    & $m_{K_{s}}$ & $L_{ir}$ & Exp. time (s)     & $k_{s,lim}$ & $m_{K_{s}}^{*}$ \\
\hline
IRAS F00235+1024 & 0.58 & 00 26 06.7 & 10 41 27.6 & nl          & 17.05   & 13.15         & 1521         & 20.68     & 16.51       \\
IRAS P09104+4109 & 0.44 & 09 13 45.4 & 40 56 28.0 & Sy2         & 15.00   & 13.24         & 2341         & 20.81     & 15.90       \\
IRAS F10026+4949 & 1.12 & 10 05 52.5 & 49 34 47.8 & Sy1         & 16.85   & 14.00         & 3511         & 21.02     & 17.95       \\
IRAS F10119+1429 & 1.55 & 10 14 37.8 & 14 15 59.7 & QSO         & 16.37   & $\sim14.31$   & 3862         & 20.87     & 18.69       \\
LBQS 1220+0939   & 0.68 & 12 23 17.9 & 09 23 07.3 & QSO         & 17.54   & $\sim13.08$   & 1170         & 20.53     & 16.97       \\
IRAS F14218+3845 & 1.21 & 14 23 55.5 & 38 31 51.3 & QSO         & 17.21   & 13.26         & 3277         & 21.15     & 18.21       \\
\hline
\end{tabular}

\medskip
Magnitudes are taken from the data presented in this paper. Infrared ($1-1000\mu$m) luminosities, given in units of 
bolometric solar luminosities, are taken from \citet{row2} and \citet{far3} and rescaled to $\Lambda = 0.7$, $\Omega_0 = 0.3$ 
and $H_0 = 70$. $k_{s,lim}$ is the faintest object detected by {\it SExtractor} for each field. $m_{K_{s}}^{*}$ was derived 
from \citet{poz} for the redshift of each object. 

\end{table*}

\section{Introduction}
Since the discovery by the {\it Infrared Astronomical Satellite} (IRAS) in 1983 of a large population of galaxies with 
significant infrared (IR) emission, substantial effort has been expended on understanding the nature of the most luminous end of this 
IR galaxy population. These sources, termed Ultraluminous Infrared Galaxies (ULIRGs) if their IR luminosity exceeds 
$10^{12}L_{\odot}$, and HLIRGs if their IR luminosity exceeds $10^{13}L_{\odot}$, are found over a very wide range in 
redshift, with most lying at $z<0.1$ \citep{soi,sau}, but with a significant number lying in the range $0.1<z<0.4$, and 
a few lying at higher redshifts. Although the consensus is now that ULIRGs and HLIRGs are powered by some combination of 
violent star formation and black hole accretion surrounded by large masses of gas and dust, the triggers for 
this activity, and how these galaxies evolve, are not known. Locally, ULIRGs are thought to be mergers between 
two or more gas-rich spiral galaxies, taking place almost exclusively in poor environments, and that a 
small number of these ULIRGs evolve into optically selected QSOs \citep{soi,lee,san2,rig,far1,bus,tac,far4}. At 
higher redshifts however, the picture is less clear. It has been suggested that a greater variety of 
galaxy formation processes may play a role amongst ULIRGs and HLIRGs at high redshift \citep{far3}, a change 
which may manifest itself in their environments. 

The Hyperluminous Infrared Galaxies (HLIRGs), which generally lie at $z\geq0.3$, have been studied extensively 
since their discovery, motivated by their extreme luminosities which make them amongst the most luminous objects 
in the Universe. The first HLIRG to be found (P09104+4109, at $z=0.44$) was a cD galaxy in the core of a rich 
cluster, identified to have extreme IR emission by \citet{klei}, with a far infrared luminosity of 
$1.5\times10^{13}h_{50}^{-2}L_{\sun}$. Then, in 1991, \citet{row0} identified F10214+4724 at $z=2.286$, with an 
apparent far infrared luminosity of $3\times10^{14}h^{-2}_{50}L_{\sun}$. Later observations revealed a large mass 
of molecular gas ($10^{11}h_{50}^{-2}M_{\sun}$ \citep{br,so3}), a Seyfert emission spectrum \citep{el}, and 
evidence for lensing with a magnification of about 10 in the infrared \citep{gra,bro,ei,gr2}. These objects 
appeared to presage a new class of infrared galaxy.

Later observations of larger samples of HLIRGs uncovered a more detailed picture. Hubble Space Telescope (HST) 
imaging \citep{far2} revealed that a wide range of morphologies are present in the HLIRG population, from merging 
systems to QSOs in apparently relaxed systems. X-ray, IR, and sub-millimetre observations showed that, in all cases, 
HLIRGs are powered by a mixture of dust-enshrouded black hole accretion and violent star formation, with inferred 
star formation rates of $\geq500M_{\odot}$yr$^{-1}$ \citep{row2,ver,far3,wil}, suggesting that HLIRGs are comprised 
of both mergers between gas-rich spiral galaxies, and young galaxies going through their maximal star formation 
epochs whilst harbouring an AGN. 

Despite this progress, the role of HLIRGs in the broader picture of galaxy and AGN evolution remains unclear. It is not 
known whether HLIRGs as a class are a simple extrapolation of the local Ultraluminous Infrared Galaxies (ULIRGs, 
$L_{ir}>10^{12}L_{\odot}$) making them mostly mergers between gas-rich spirals, or whether a wider range of 
galaxy formation processes play a role in HLIRG evolution. Also, the links between HLIRGs and QSOs 
at comparable redshifts are not well understood. Locally, it is thought that some fraction of ULIRGs evolve into 
optically selected QSOs \citep{san1,far1,tac}, but it is not known whether this is also true in the distant Universe.

Many of these unknowns result from two major obstacles in studying HLIRG evolution. Firstly, HLIRGs contain very large 
masses of gas and dust, making observations of the galaxies themselves at all wavelengths (except perhaps the far-infrared and sub-millimetre) 
prone to obscuration bias. Secondly, the presence of a luminous starburst and AGN in all HLIRGs means that observations 
will be affected by the orientation of the HLIRG relative to us. These problems can however be partly overcome by 
examining the environments of HLIRGs. Since the determination of environments is independent of orientation and dust content, they are a useful 
tool in studying AGN evolution, and have been used extensively in studying both normal and active galaxies 
\citep{los,yee0,hil,lov,wol1,wol2,mcl,sng}. Studying the environments of HLIRGs therefore can help clarify the 
relations between HLIRGs and other AGN classes. 

In this paper, we investigate the environments of six HLIRGs, using deep wide 
field $K_{s}$-band imaging. Observations are described in \S2 and analysis is described in \S3. 
Results are presented in \S4, with discussion in \S5. Finally, our conclusions are summarized in \S6. 
Unless otherwise stated, we assume $\Lambda = 0.7$, $\Omega_0 = 0.3$ and $H_0 = 70$ km s$^{-1}$ Mpc$^{-1}$.

\section{Observations}

We selected for observation six HLIRGs from the sample presented by \citet{row2}. The sample, their redshifts 
and other basic data are presented in Table \ref{hlirgsample}. Five of these objects were selected to lie 
approximately in the redshift range $0.6 < z < 1.6$, where the greatest evolution in the IR galaxy population 
is thought to occur (e.g. \citet{row1}). Additionally, we observed one further HLIRG, P09104+4109, which lies at $z=0.44$ and is 
already known to lie in a rich cluster, to act as a control for our observation and analysis methods, though we 
do not include this source in the discussion. None of our targets show any evidence for significant gravitational 
lensing. 

Observations were made on 25-26th December 2001 using the INGRID wide field near-infrared imager and a $K_{s}$-band 
filter, on the 4.2m William Hershel Telescope (WHT).  INGRID is a $1024\times1024$ pixel array, with a scale of 
0.238$\arcsec$ pix$^{-1}$, corresponding to a field-of-view of $\sim17\arcmin^{2}$. At the redshifts of our 
sample this corresponds to a physical field-of-view of $>1.36$Mpc. The targets were centred approximately in the 
INGRID field-of-view, with exposure times selected to reach a minimum depth of $M_{K_{s}}^{*} +2$ at the redshift 
of each object. Observing conditions were generally good, with little cloud cover and seeing of $\sim0.8\arcsec$, 
however the atmospheric stability was variable, resulting in non-photometric nights. The total exposure time for 
each object was divided into several nine point `box' dither patterns, with a $16.7\arcsec$ offset between each 
position, to allow the subtraction of cosmic rays, hot pixels, and the infrared sky background. As the INGRID 
field of view is too small to reliably estimate field galaxy counts from the edges of the HLIRG fields, separate 
control fields were also observed, with similar galactic latitudes and exposure times as the sample. For 
photometric calibration we observed a selection of infrared standard stars throughout each night, at several 
different airmasses.

\section{Data reduction}

Following debiasing and flatfielding, the data were reduced using our own custom-written 
{\footnotesize IRAF} pipeline, based in part on the {\it Quicklook} INGRID data reduction 
pipeline from the Isaac Newton Group (ING). As the exposure time at each position in the dither pattern 
was only 30 seconds, particular care was taken in accounting for bright sources when 
subtracting sky noise. Source masks were created for the individual frames in each dither pattern 
by first creating an approximate estimate of the sky by median combining the nine frames in 
each dither pattern without applying the dither offsets. This `dummy' sky frame was then 
subtracted from each frame in the dither pattern to reveal the brightest sources, which were then 
masked out when creating the `real' sky frame. The nine frames in each dither pattern were 
combined using the {\footnotesize IRAF} task {\footnotesize IMCOMBINE}, with dither offsets 
calculated by centroiding two or more bright sources common to each frame. After each dither 
pattern was combined into a single image, sky subtraction was performed using the `real' sky 
frames and pixel masks described earlier. It was found that this sky subtraction method worked 
well in dealing with variations in the infrared sky between different dither patterns, over 
the long exposure time for each object. The combined images from each dither pattern were then 
stacked together to produce a final image $I$:

\begin{equation}
I = {{\sum_i\left(I_i\sigma^{-2}_i\right)}\over{\sum_i\sigma^{-2}_i}}
\end{equation}

\noindent where $I_{i}$ and $\sigma_{i}$ are the individual images from each dither pattern, and 
their standard deviations, respectively. The standard deviations were derived from the noise fluctuations 
in each image. Noisy edges, where the total exposure times were shorter due to the dither pattern, were 
clipped off. The resulting field of view, at $13 \arcmin \times 13 \arcmin$, was still sufficient to quantify 
the environments of the sample. The final images were of excellent quality, and flat 
to better than $1\%$ across the width of the frame, and were in all cases of much higher quality 
than the images from the {\it Quicklook} reduction. Our reduction pipeline, {\footnotesize `INREP'}, 
is available for general use via the Isaac Newton Group (ING) web pages \footnote{http://ing.iac.es/Astronomy/Ingrid}.

Sources were extracted and catalogued using the {\it SExtractor} package \citep{ber}. For source 
extraction, we adopted the conservative criterion that a source constitutes at least four contiguous 
pixels, with a significance of detection of at least $3\sigma$ above the background. The default {\it SExtractor} extraction 
filter was used to detect faint extended objects, with 32 de-blending thresholds, a cleaning efficiency 
of 1 and a contrast parameter of 0.005. The background estimation was mesh based, with a mesh size of 
64 and a filter size of 3. As many of the sources in the frames were faint and slightly extended, the 
magnitudes were calculated using corrected isophotal (ISOCOR) magnitudes within {\it SExtractor}. The 
calibration zeropoint magnitudes for each object frame and control frame were calculated using the 
observations of the infrared standard stars. The errors on the final magnitudes arising from the standard 
star calibrations are $\Delta m = 0.05$, with a further error of $\Delta m = 0.05$ from the source extraction. 
To correct our measured magnitudes to the rest-frame $K_{s}$-band, we computed $k$-corrections assuming a 
standard power law, with $\alpha=2$.

\section{Analysis}\label{analyse}

To evaluate the environmental richness of each object in our sample we use two independent clustering 
measures; the $B_{\rmn{gg}}$ galaxy-galaxy correlation statistic \citep{los}, which for our sample we refer to 
as the $B_{\rmn{gh}}$ galaxy-HLIRG correlation statistic, and the $N_{0.5}$ statistic \citep{hil}. The 
$B_{gh}$ statistic is the amplitude of the spatial cross-correlation function, and relies on knowing the volume 
density of galaxies around the HLIRG, which itself requires knowledge of the luminosity function (LF). 
The $N_{0.5}$ statistic is a simple, more direct counting statistic, and involves counting all the sources 
within 0.5Mpc of the target that lie within a certain magnitude range. Although the $B_{\rmn{gh}}$ statistic 
is physically more meaningful than the $N_{0.5}$ statistic, the near-infrared luminosity function is not 
known with great precision at high redshifts. Therefore, we have adopted the $B_{\rmn{gh}}$ statistic as our primary 
measure of environmental richness, but we also use the $N_{0.5}$ statistic as an independent check. 

\begin{table*}
\centering
\caption{HLIRG clustering statistics \label{clusterstats}}
\begin{tabular}{@{}lccccccc}
\hline 
Name               & $N_{0.5}$ & $\sigma_{N_{0.5}}$ & $A_{gh}      $                 & $\sigma_{A_{gh}}$  & $B_{gh}$     & $\sigma_{B_{gh}}$ & Abell class \\
                   &           &                    & $\times10^{3}\rmn{rad}^{0.77}$ &                    & Mpc$^{1.77}$ &                   &             \\
\hline
IRAS F00235+1024   & 7.69      & 2.77               & 0.66                           &   0.28             & 208.63       &  88.31            & field       \\
IRAS P09104+4109   & 29.65     & 5.45               & 2.75                           &   0.40             & 819.75       &  119.77           & 2           \\
IRAS F10026+4949   & 17.81     & 4.22               & 1.73                           &   0.40             & 611.46       &  141.01           & 1           \\
IRAS F10119+1429   & 8.88      & 2.98               & 0.72                           &   0.31             & 384.32       &  167.33           & 0           \\
LBQS 1220+0939     & 3.77      & 1.94               & 0.40                           &   0.26             & 143.78       &  94.25            & field       \\
IRAS F14218+3845   & 4.45      & 2.11               & 0.06                           &   0.21             & 21.84        &  81.27            & field       \\
\hline
\end{tabular}

\medskip
Quoted errors only include the counting errors. The methods used to calculate the $B_{gh}$ and $N_{0.5}$ statistics, and the Abell 
classes, are described in \S \ref{analyse} and \S \ref{resoolt} respectively.

\end{table*}

To compute the $B_{\rmn{gh}}$ statistic, we follow the prescription of \citet{los}. As this statistic is 
not trivial to evaluate, we provide a detailed outline of the procedure here. The spatial cross-correlation 
function, $\xi(r)$, can be described as a simple power law:

\begin{equation}
\xi(r) = B_{\rmn{gh}}r^{-\gamma}
\end{equation}

\noindent and is derived from the galaxy-HLIRG angular cross-correlation function $w(\theta)$:

\begin{equation}
w(\theta) = A_{\rmn{gh}} \theta^{1-\gamma}
\end{equation}

\noindent by deprojecting 
from the celestial sphere. The amplitude of the angular cross-correlation function, $A_{\rmn{gh}}$, is 
essentially the richness measure of the number of galaxies about a point on the sky, and is converted to 
$B_{\rmn{gh}}$, the richness measure of the number of galaxies about a point in {\it space}, by translating 
the angular cross-correlation function $w(\theta)$ to the spatial cross-correlation function $\xi(r)$.

To perform this conversion, the $A_{\rmn{gh}}$ statistic is evaluated:

\begin{equation}
A_{\rmn{gh}} = {{N_T - N_B}\over{N_B}}{{3-\gamma}\over{2}}\theta^{\gamma-1}
\end{equation}

\noindent where $N_T$ is the total number of galaxies within a radius
$\theta$ around the target, corresponding to 0.5 Mpc at the target
redshift, and $N_B$ is the number of galaxies within the same radius in the 
control field for that object. We make the assumption \citep{wol1,wol2}
that $\gamma = 1.77$. The precise choice of $\gamma$ will however not affect 
the derived clustering amplitudes, as it has previously been shown \citep{pre1,pre2}
that $B_{\rmn{gh}}$ is insensitive to the choice of $\gamma$, as
long as $\gamma \sim 2$.

The $B_{\rmn{gh}}$ statistic can now be calculated, and normalised
to the integral LF, $\Phi(m_l,z)$ which represents the number of
galaxies more luminous than $m_l$ per unit co-moving volume at
redshift $z$:

\begin{equation} B_{\rmn{gh}} =
{{\rho_gA_{\rmn{gh}}}\over{\Phi(m_l,z)I_\gamma}}d^{\gamma-3}_{\theta}
\end{equation}

\noindent where $\rho_g$ is the average surface density of background
galaxies, $d_{\theta}$ is the angular diameter distance to the
target and $I_{\gamma}$ is an integration constant:

\begin{equation}
I_\gamma = \frac{2^\gamma}{\gamma
-1}\frac{\Gamma^2[(\gamma+1)/2]}{\Gamma(\gamma)} \sim 3.78
\end{equation}

\noindent The integration to $\Phi(m_l,z)$ is performed by taking a
universal LF of the form \citep{sch}:

\begin{equation}
\phi(L) \sim
\left(\phi^*/L^*\right)\left(L/L^*\right)^{\alpha}\exp\left(-L/L^*\right)
\end{equation}

\noindent and integrating down to the completeness limit at the target
redshift $L(m_l,z)$:

\begin{equation} \Phi(m_l,z) =
\int_{L(m_l,z)}^{\infty} \phi(L)dL
\end{equation}

\noindent For the parameters used in the LF, we use the most recent determination of the 
high redshift near-IR LFs as given by \citet{poz}, who have determined 
the evolution of the near-IR LF in the $J$- and $K_{s}$-bands in redshift bins of $z =
\left<0.20,0.65\right>$ and $z = \left<0.75,1.30\right>$, using a spectroscopic survey 
of a magnitude limited sample of galaxies with $K_s<20$. In terms of absolute magnitudes 
the luminosity function can be expressed as:

\begin{equation}
\phi(M) = Cf(M)^{(\alpha+1)}\exp(-f(M))
\end{equation}

\noindent where the constant $C = 0.4\ln(10)\phi^*$, and: 

\begin{equation}
f(M) = 10^{(-0.4(M-M^*))} 
\end{equation}

\noindent As some of the galaxies in our sample lie outside of the redshift bins given 
by \citet{poz}, we must make assumptions as to the nature of the LF at these magnitudes. In 
the redshift range $0.2<z<0.65$, we have assumed $\alpha=-1.25$,  $M^*=-25.64$ and 
$\phi^*=6.11\times10^{-4}$Mpc$^{-3}$. In the redshift range $0.75<z<1.55$ we have assumed 
$\alpha=-0.98$,  $M^*=-25.54$ and $\phi^*=9.98\times10^{-4}$Mpc$^{-3}$. We note that \citet{poz} 
find only mild evolution in the $K_{s}$-band luminosity function over the redshift range 
$0<z<1.3$, hence our use of their luminosity function at $z=1.55$ is unlikely to be a major 
source of error. 

The errors in $A_{\rmn{gh}}$ and $B_{\rmn{gh}}$ were
calculated following the prescription given by \citet{yee}:

\begin{equation}
{{\Delta A_{\rmn{gh}}}\over{A_{\rmn{gh}}}}={{\Delta
B_{\rmn{gh}}}\over{B_{\rmn{gh}}}} = { {\sqrt{(N_T - N_B) +
1.3^2N_B}}\over{N_T - N_B}}
\end{equation}

\noindent which is based only upon (non-Poissonian) counting statistics. Other systematic 
errors are discussed in \S\ref{errbudg}.

\begin{figure*}
\begin{minipage}{180mm}
\epsfig{figure=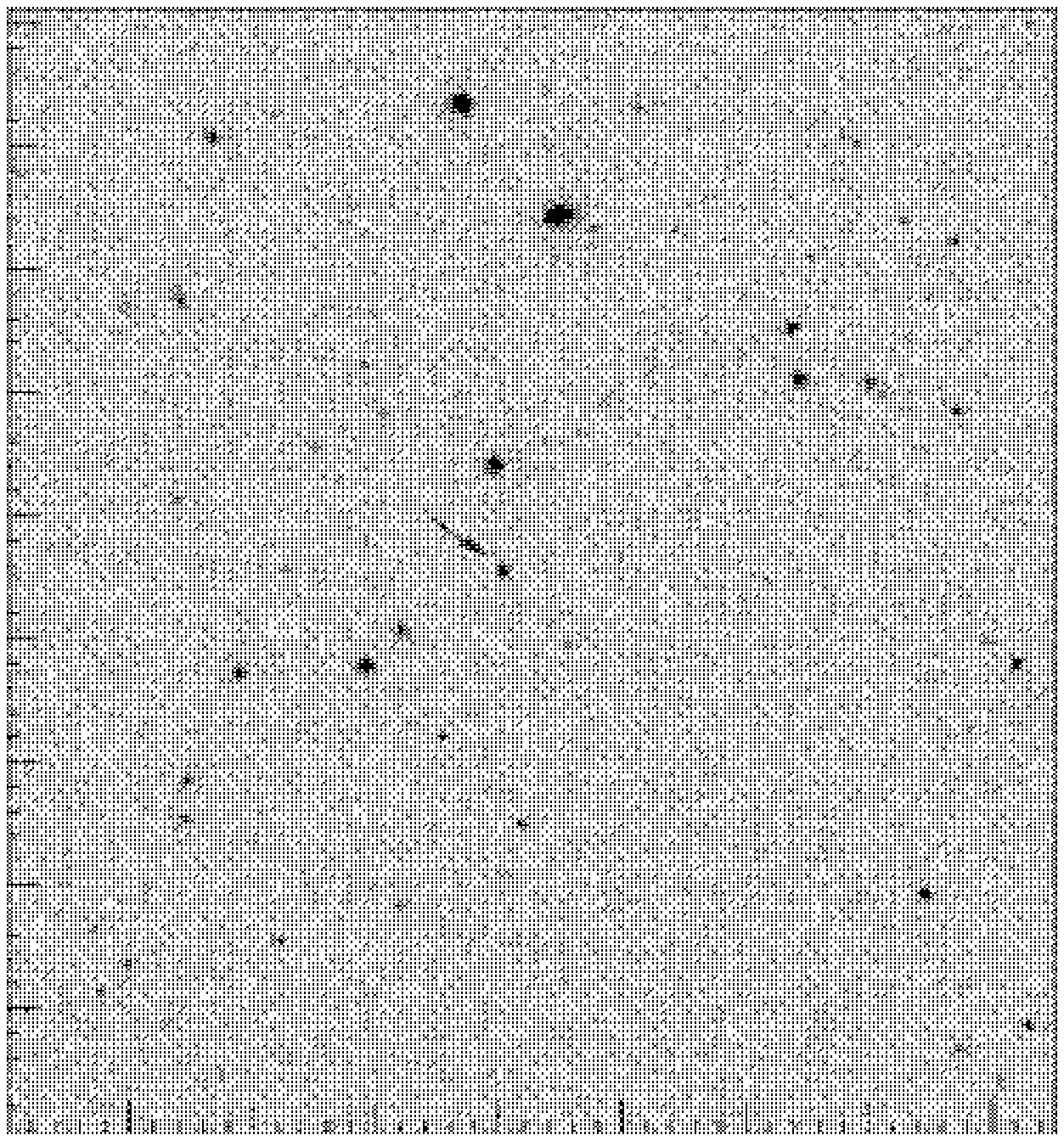,width=68mm}
\epsfig{figure=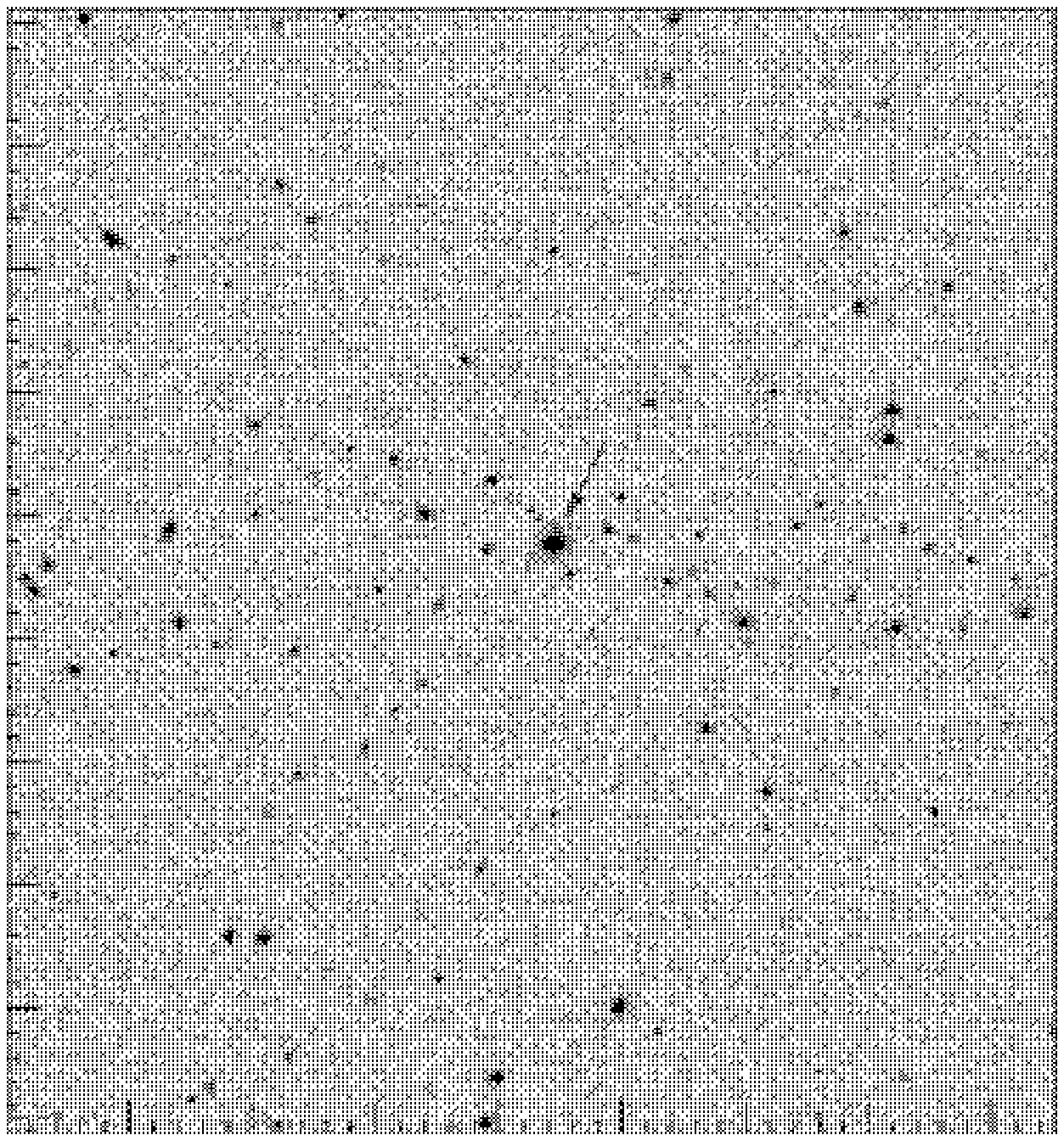,width=68mm}
\end{minipage}
\begin{minipage}{180mm}
\epsfig{figure=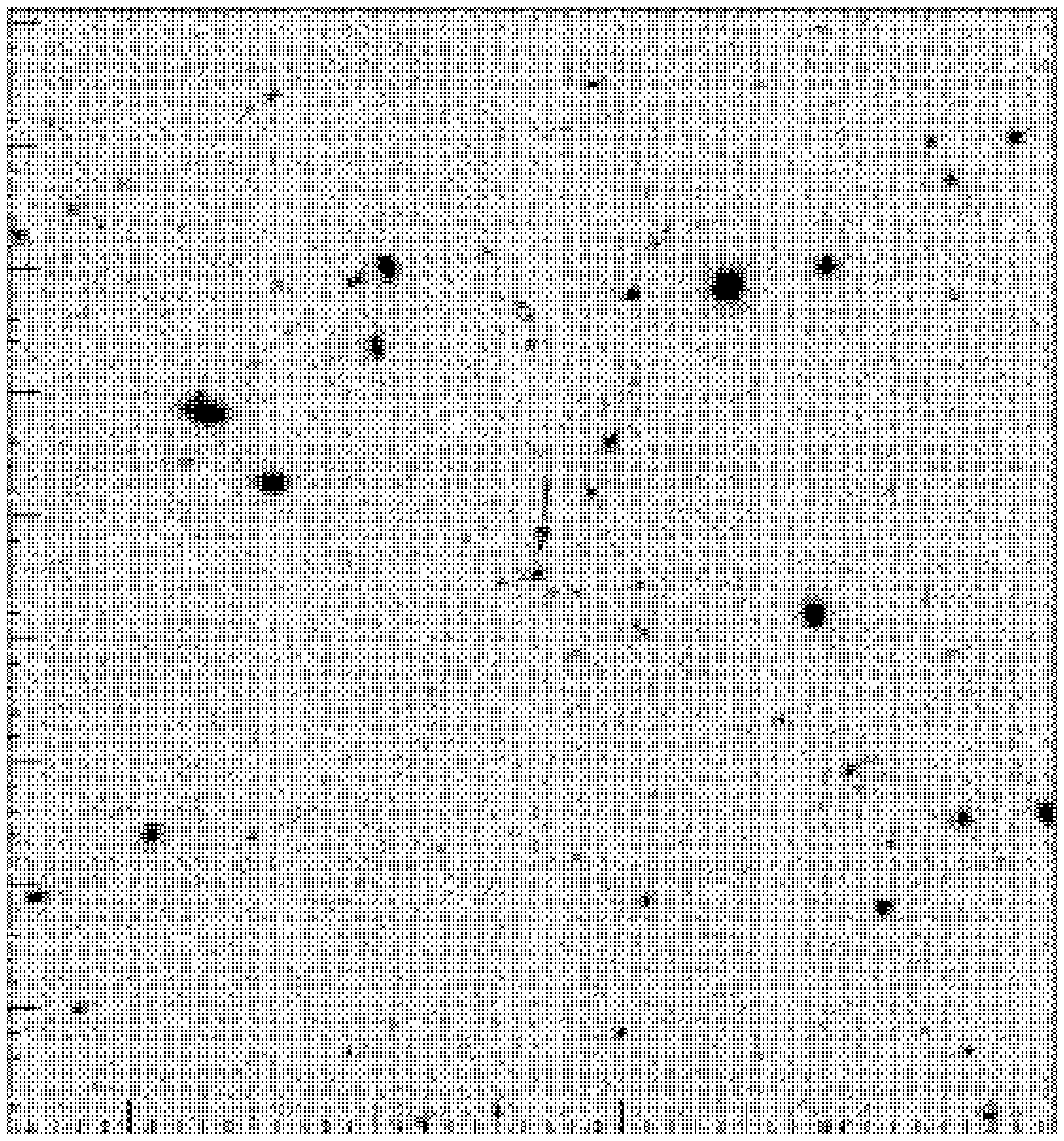,width=68mm}
\epsfig{figure=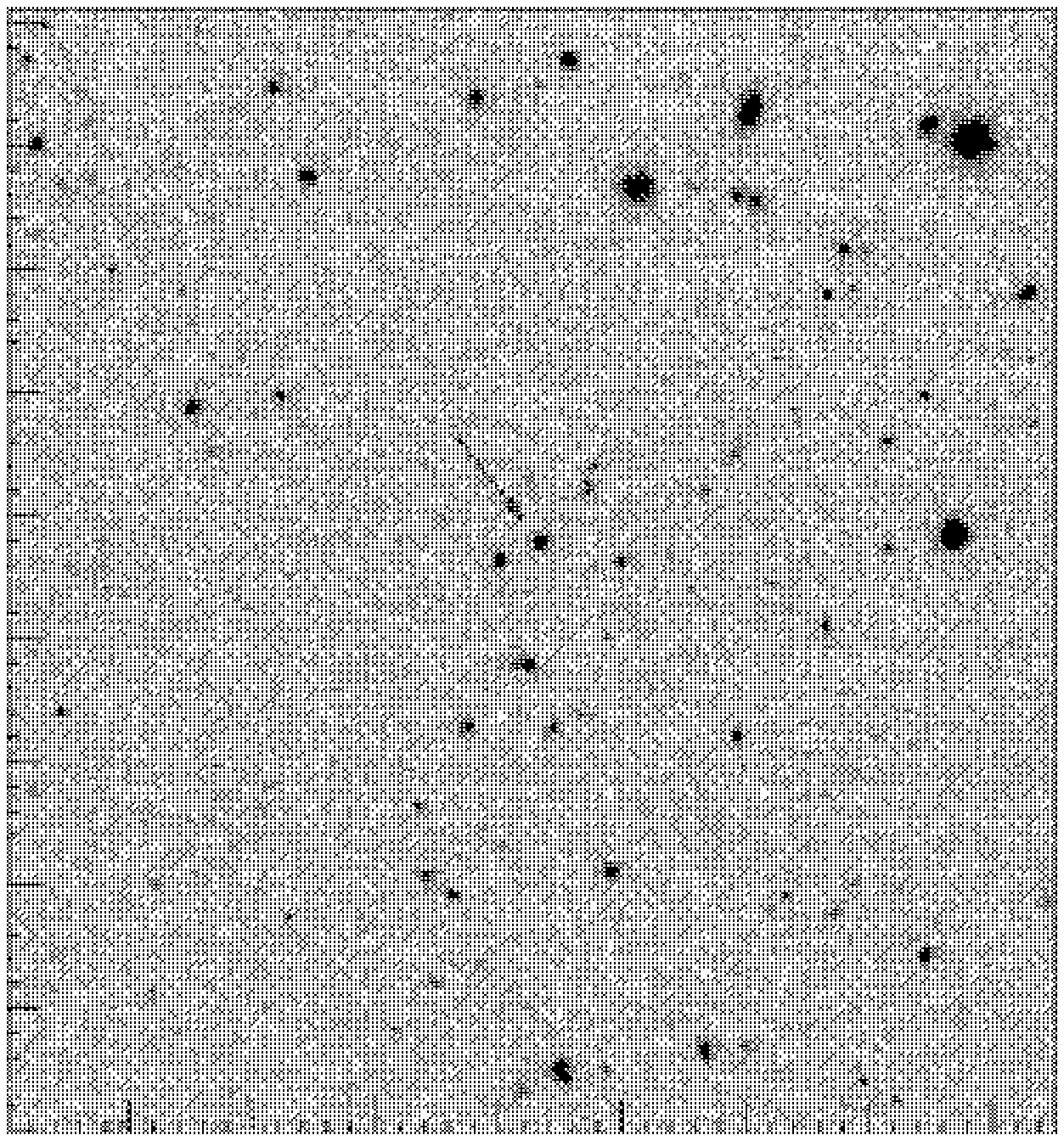,width=68mm}
\end{minipage}
\begin{minipage}{180mm}
\epsfig{figure=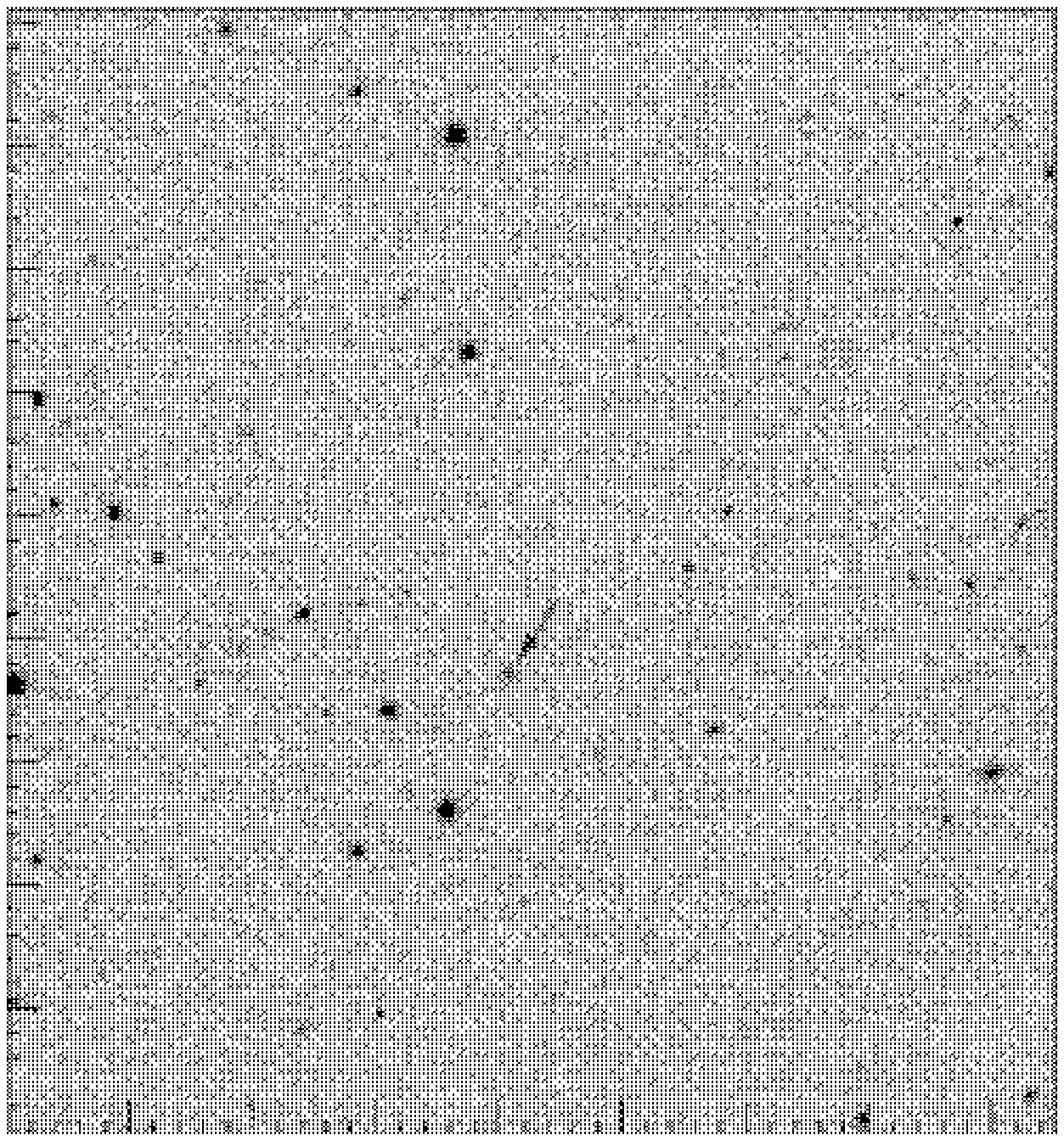,width=68mm}
\epsfig{figure=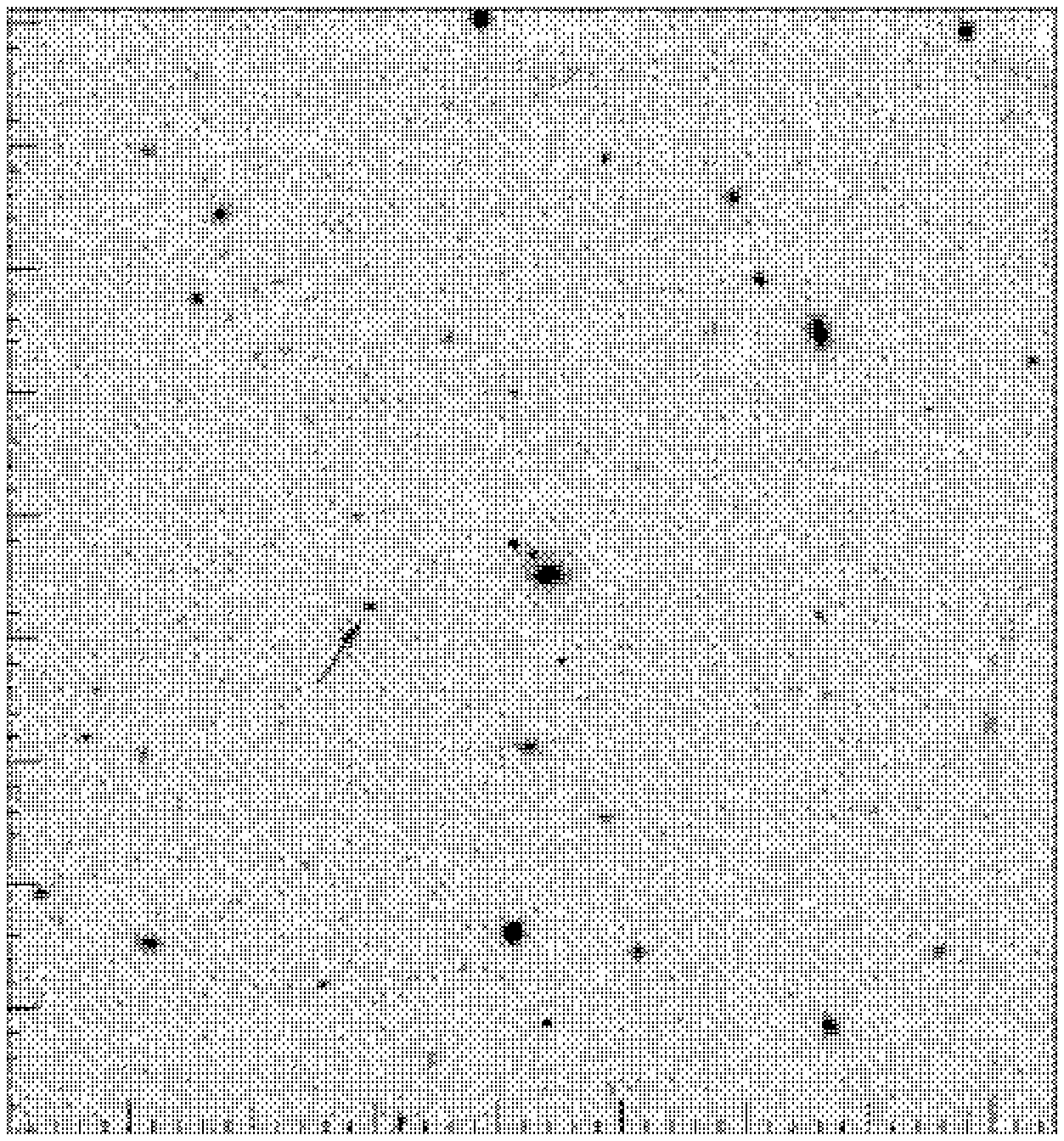,width=68mm}
\end{minipage}
\caption{
$K_{s}$-band images of the fields around each HLIRG. An arrow indicates the HLIRG. Large tick marks correspond to $25\arcsec$. 
L-R: (top row)  F00235+1024 and P09104+4109, (middle row) F10026+4949 and F10119+1429, (bottom row) LBQS1220+0939 and F14218+3845. 
\label{gal_images}}
\end{figure*}

\begin{figure*}
\begin{minipage}{180mm}
\epsfig{figure=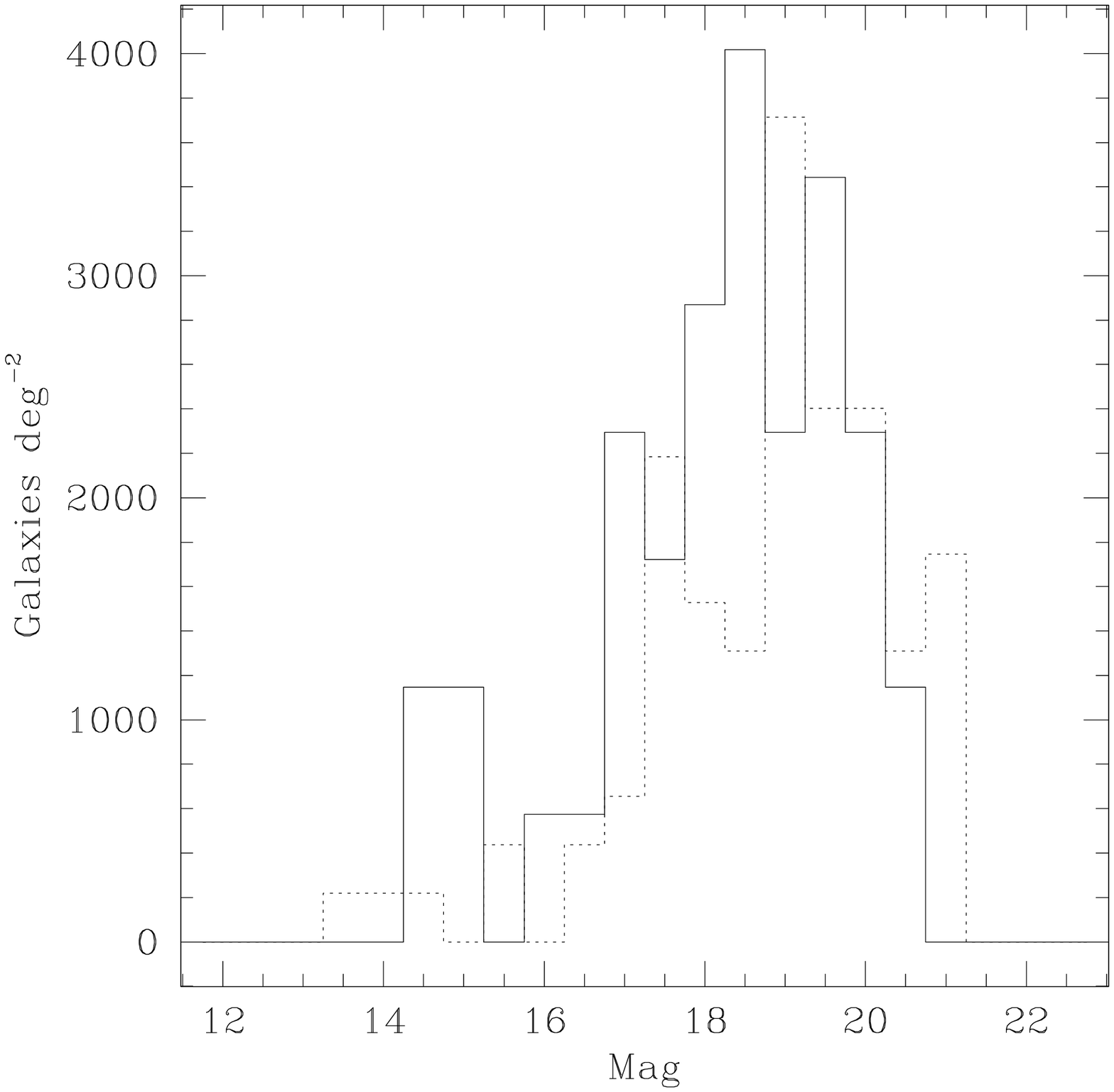,width=73mm}
\epsfig{figure=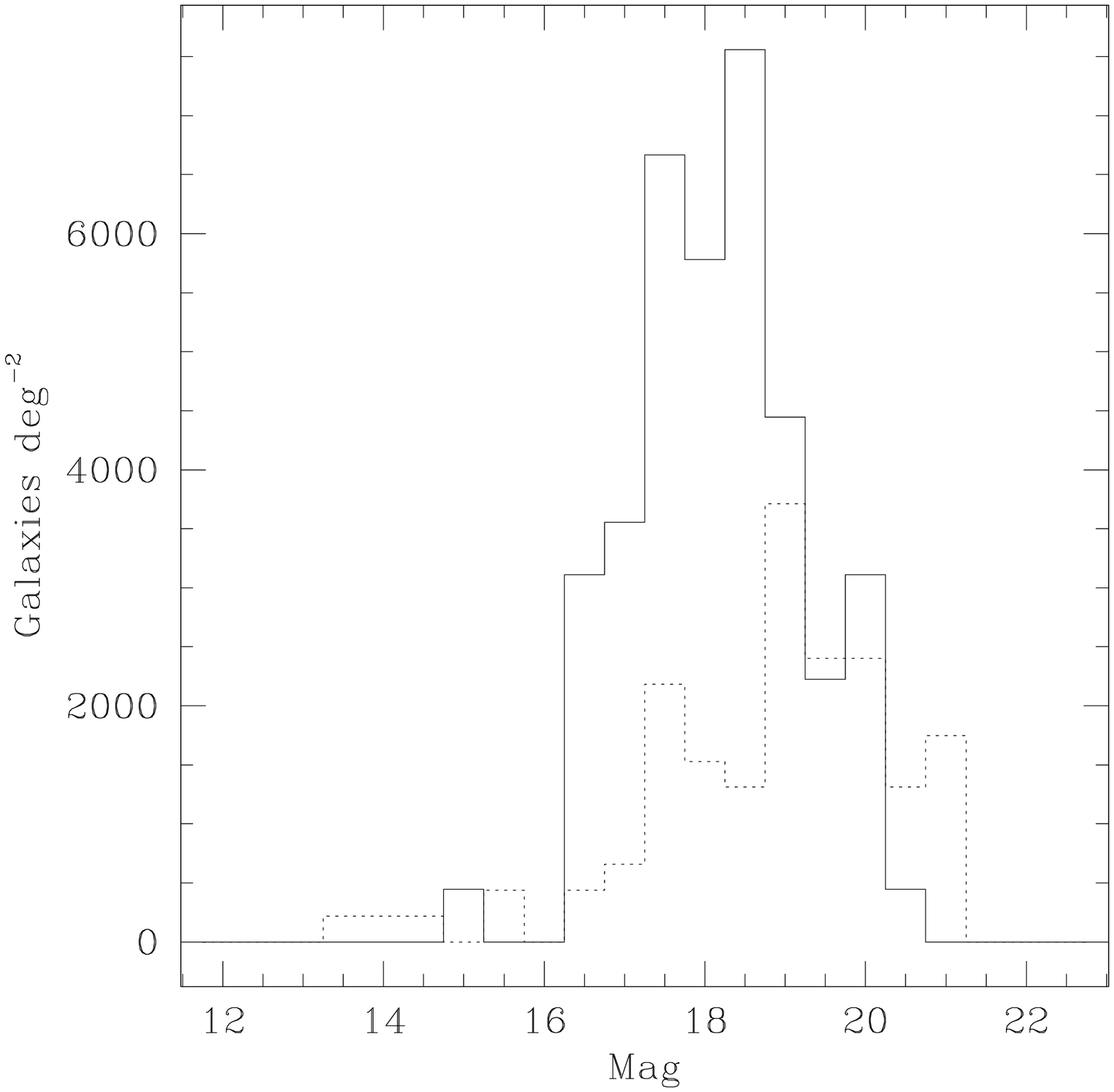,width=73mm}
\end{minipage}
\begin{minipage}{180mm}
\epsfig{figure=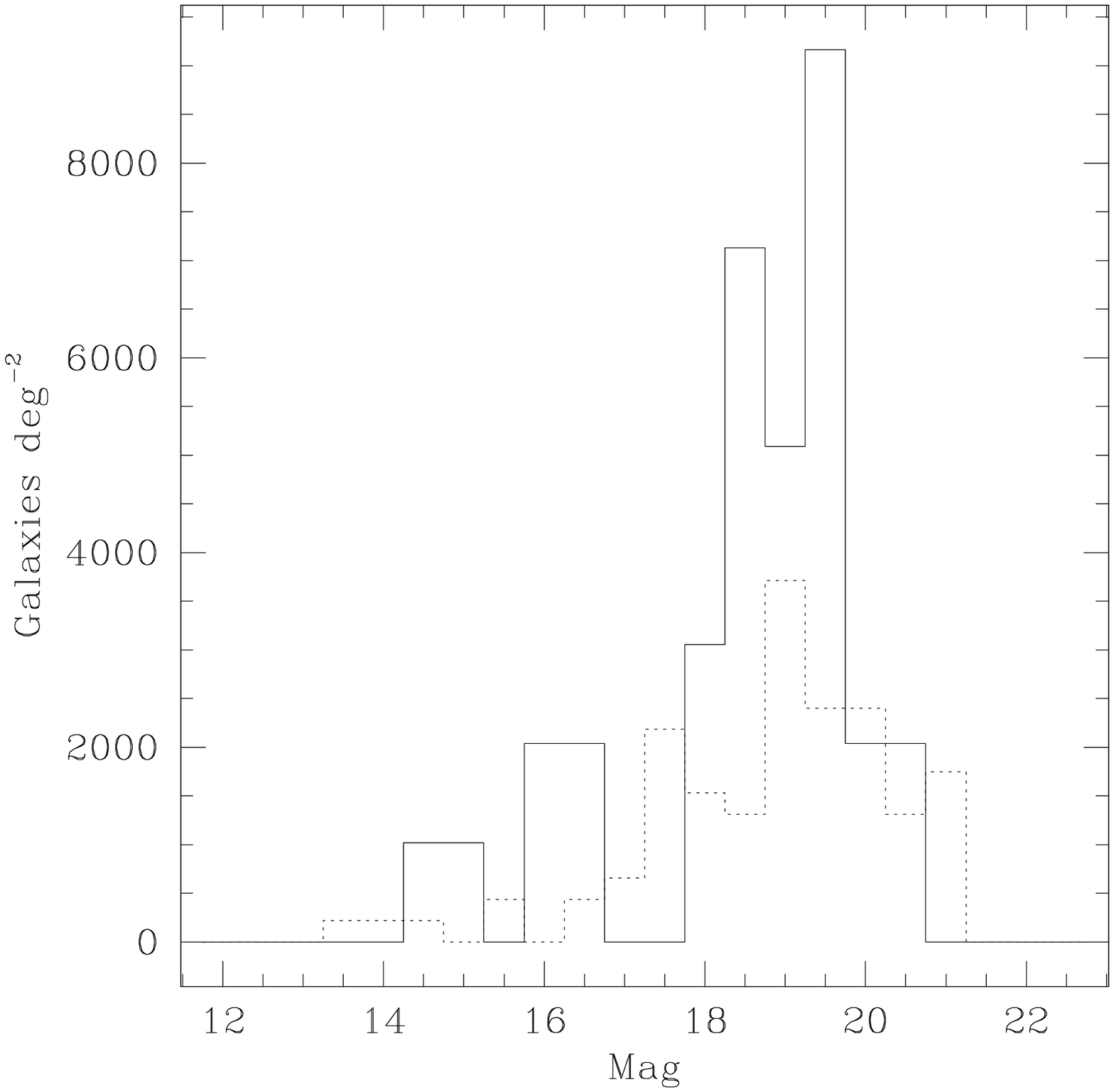,width=73mm}
\epsfig{figure=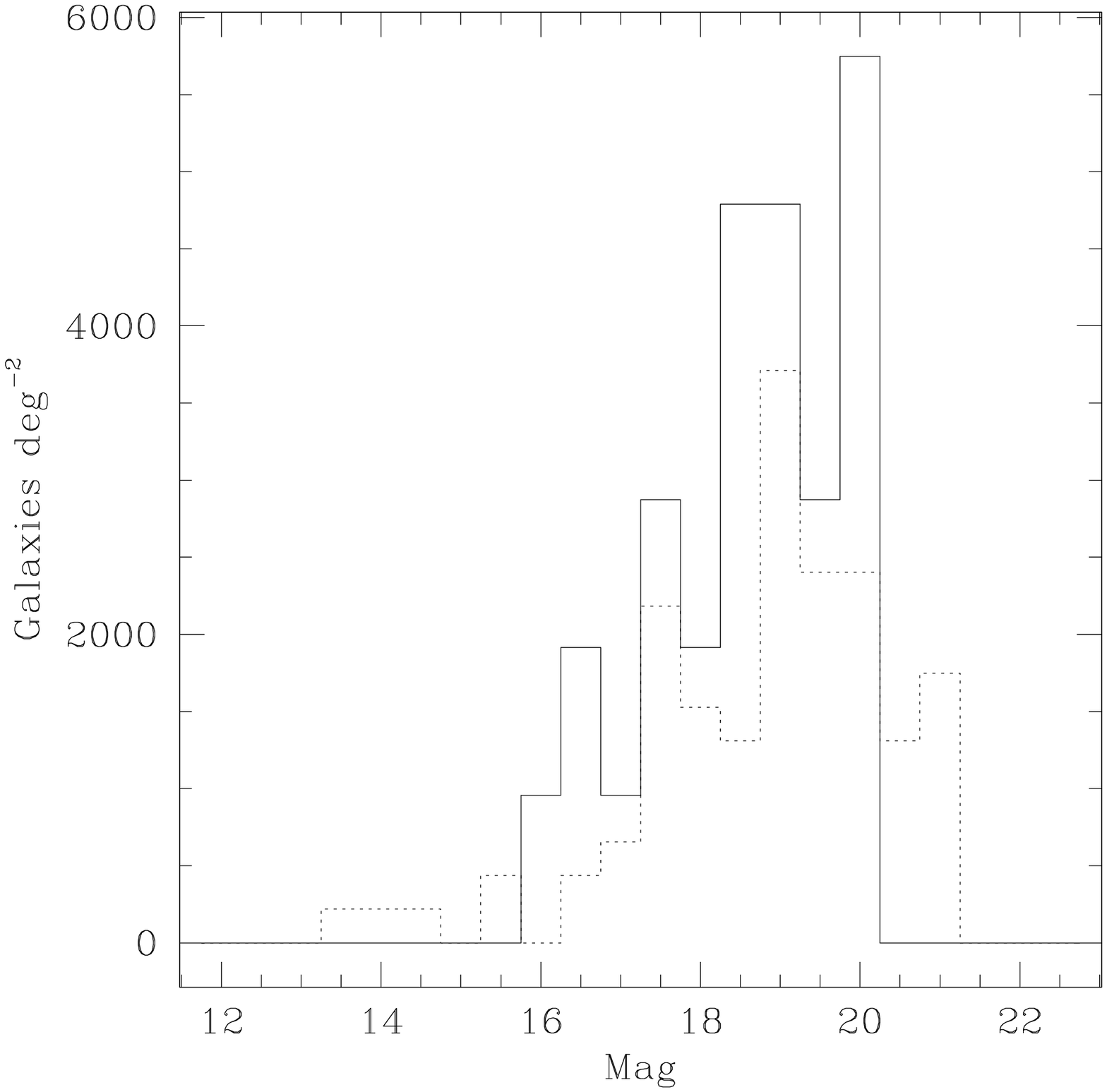,width=73mm}
\end{minipage}
\begin{minipage}{180mm}
\epsfig{figure=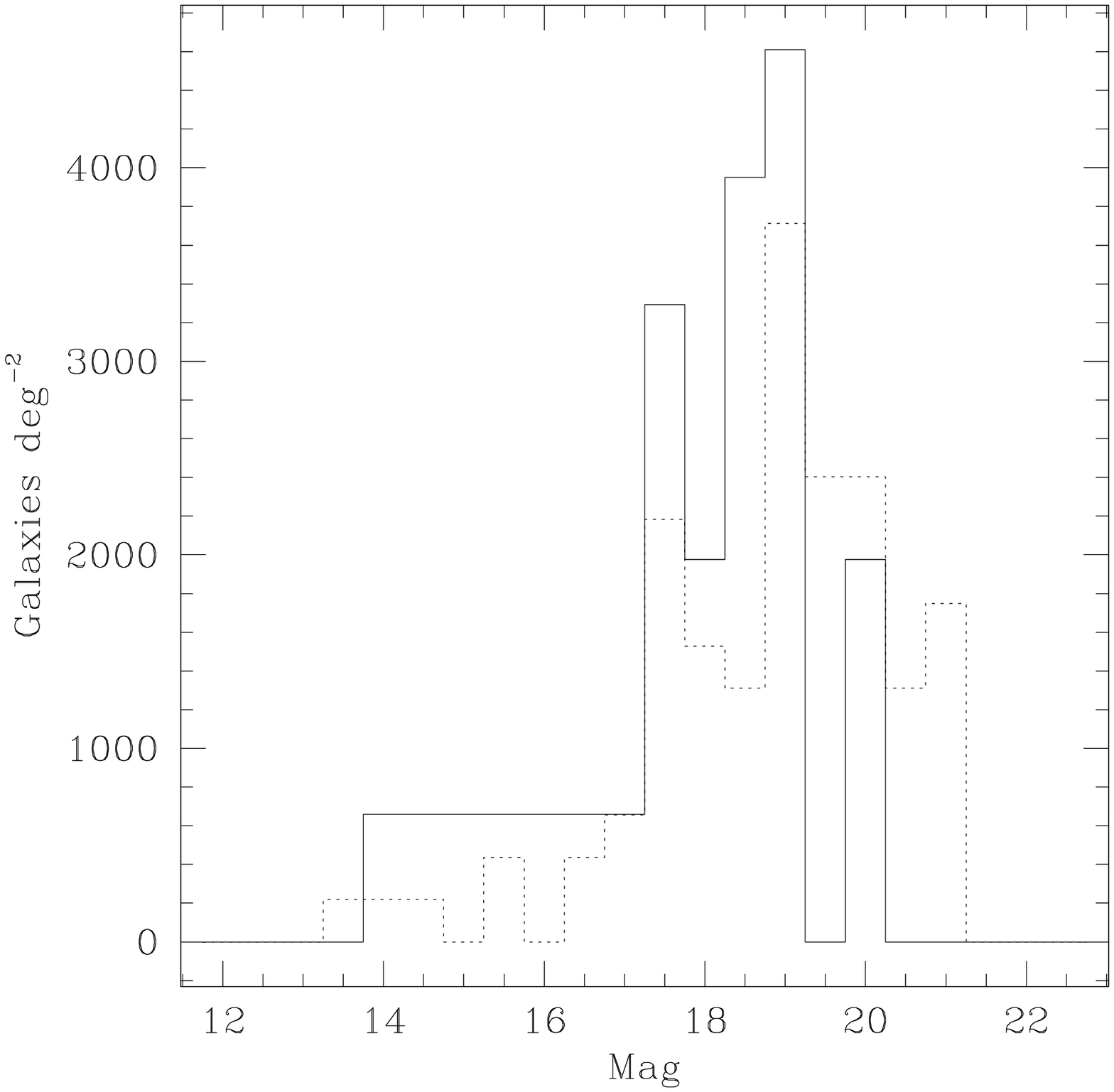,width=73mm}
\epsfig{figure=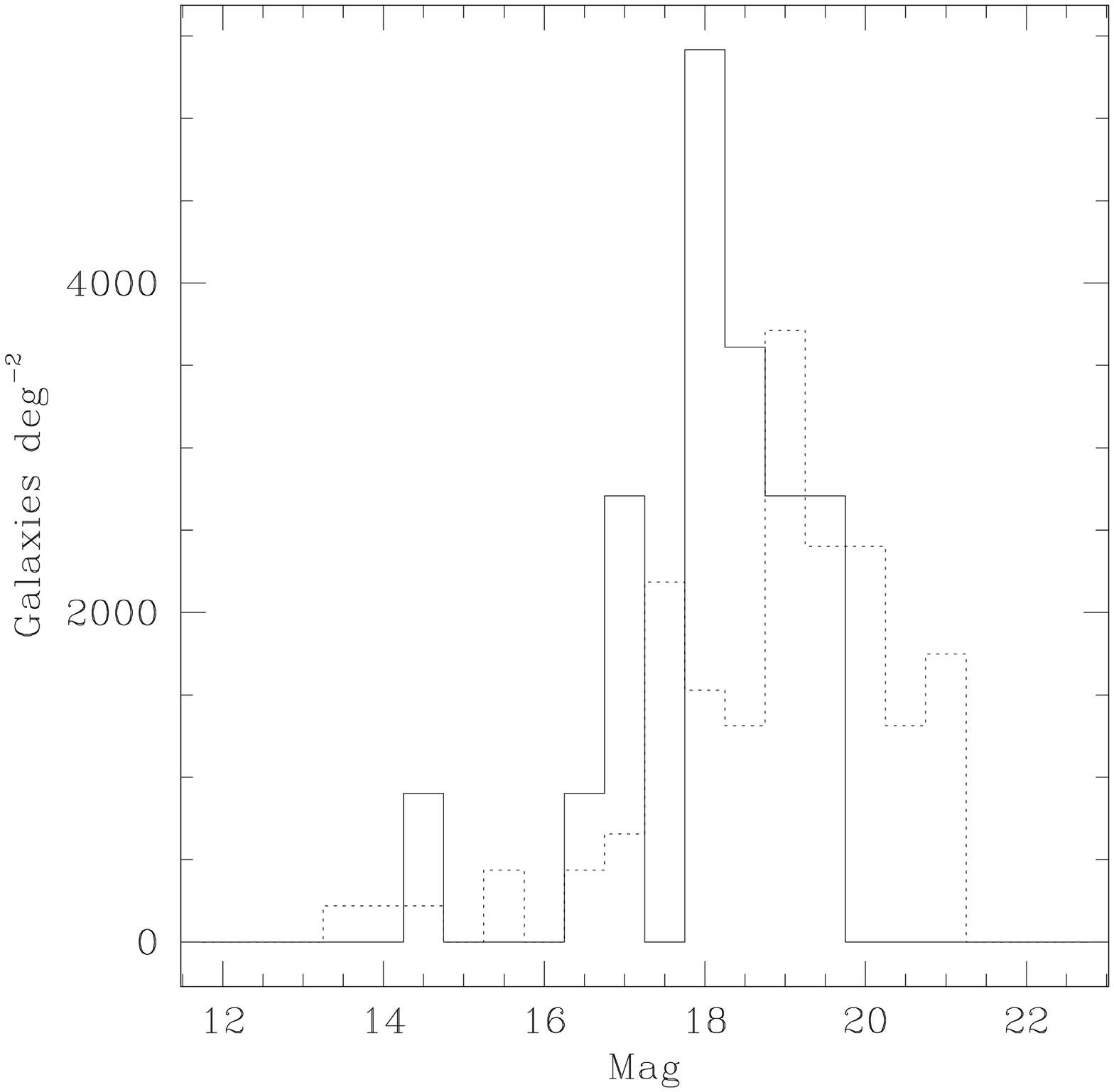,width=73mm}
\end{minipage}
\caption{
Galaxy counts in a circular region 0.5Mpc in radius around each HLIRG. The solid line shows the counts in the field around the 
HLIRG, and the dashed line shows the counts in the control field. L-R: (top row)  F00235+1024 and P09104+4109, (middle row) 
F10026+4949 and F10119+1429, (bottom row) LBQS1220+0939 and F14218+3845. Error bars have been omitted for clarity. 
 \label{gal_excess}}
\end{figure*}

To compute the $N_{0.5}$ statistic, we follow the procedure described by \citet{hil}. 
This statistic is computed by counting all sources within a 0.5
Mpc radius of the target, with a magnitude in the interval $\left<m,m+3\right>$, where 
$m$ is the magnitude of the target. Subtracted from this number is the expected number 
of background field galaxies in the same magnitude interval calculated from the control 
fields. The resulting number is the $N_{0.5}$ statistic. This measurement does not require 
integration of a luminosity function, although if the luminosities of HLIRGs and cluster 
galaxies evolve very differently with redshift then the $N_{0.5}$ statistic will give 
inaccurate results. There is no evidence why this should be a major effect however, 
so we do not take it into account. From Table \ref{hlirgsample}, the limiting magnitude 
for each object is fainter than $m_{HLIRG}+3$ except for two cases; the image for F10119+1429 
reaches $m_{HLIRG}+2.18$, and the image for F14218+3845 reaches $m_{HLIRG}+2.94$. In both 
these cases however the limiting depth is sufficiently near $m_{HLIRG}+3$ that the $N_{0.5}$ 
statistic is still usable.

\section{Results} \label{resoolt}

\subsection{Clustering statistics}

Images of the fields around each HLIRG are presented in Figure \ref{gal_images}. In Figure \ref{gal_excess} we plot the galaxy counts in a 
0.5Mpc region around each HLIRG, and the counts for the associated control field. The $B_{gh}$ and $N_{0.5}$ statistics for each object, 
and their errors, are given in Table \ref{clusterstats}. We also quote the approximate Abell classes of our sample 
in this table, based on the conversions given by \citet{hil}, in our cosmology. We note however that these conversions are arbitrary, and 
hence we base our quoted Abell classes on both the $B_{gh}$ and $N_{0.5}$ statistics. 

A wide variety of environments can be seen amongst the objects in the sample. P09104+4109 resides in an Abell 2 cluster, in agreement with 
previous results \citep{hin}. For the remaining objects, three reside in poor environments, and two, F10026+4949 and F10119+1429, 
reside in clusters, of Abell class 1 and 0 respectively, though we note that the detection of clustering for F10119+1429 is marginal, 
at $\sim2.5\sigma$. Excluding P09104+4109, the error weighted mean clustering amplitude for the remaining 5 objects is 
$\langle B_{gh} \rangle=190\pm 45$Mpc$^{1.77}$. 

\begin{figure}
\rotatebox{0}{
\centering{
\scalebox{0.41}{
\includegraphics*[18,144][592,718]{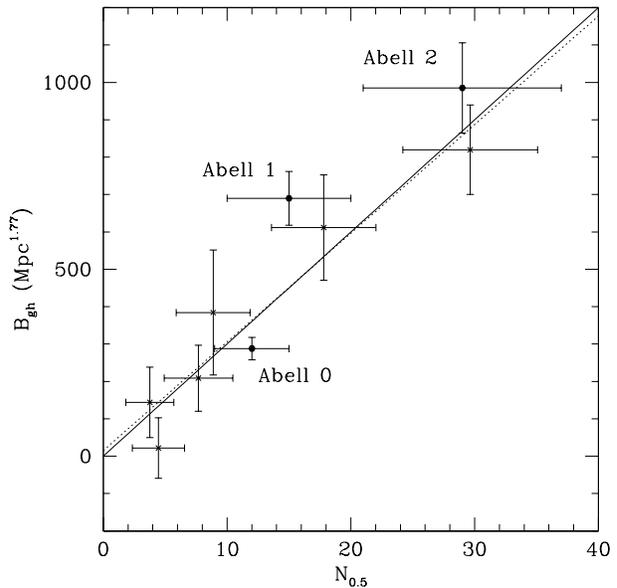}
}}}
\caption
{
$B_{gh}$ vs. $N_{0.5}$ for our sample. Filled hexagons are the Abell calibrations 
taken from \citet{hil}, stars are our data. The solid line shows the relation
$B_{gh}\propto30N_{0.5}$ found by \citet{hil}. The dashed line is a least squares 
fit of our data, which gives $B_{gh}=29.07N_{0.5}+14.95$
\label{bggn05}
}
\end{figure}

We can compare the values of $N_{0.5}$ and $B_{gh}$ to each other, as these quantities have 
been used by many previous authors, and a well defined relation has been found between them. \citet{hil} derive $B_{gh}\propto30N_{0.5}$. 
In Figure \ref{bggn05} we plot this relation, together with $B_{gh}$ vs. $N_{0.5}$ for the objects in our sample, and the conversions 
between $B_{gh}$ and Abell class given by \citet{hil}. We also plot the best fit linear relation between $B_{gh}$ and $N_{0.5}$ for our data. 
The best fit is well matched to the relation derived by \citet{hil} and the objects in the sample follow this relation closely, 
hence we conclude that our computed values of $N_{0.5}$ and $B_{gh}$ are reasonable.

\subsection{Error budget} \label{errbudg}

The errors quoted in Table \ref{clusterstats} for the $B_{gh}$ and $N_{0.5}$ statistics are only the counting errors, and do not 
include three further, potentially important sources of error. In this section, we discuss these three error sources in turn.

The first of these is the luminosity function assumed in calculating the $B_{gh}$ statistic. Due to the uncertainties in the high 
redshift $K_{s}$-band LF used in this paper, this is an important issue to address. \citet{yee} have examined the sensitivity of 
$B_{\rmn{gh}}$ to the form of the LF, and find that an error in the assumed value of $M^*$ of up to $\pm0^{\rmn{m}}.5 $ will still 
yield essentially the same results. Similarly, an error in $\alpha$ of up to $\pm 0.3$ will only affect $B_{\rmn{gh}}$ by at most 
$\sim20\%$. Although the K band LF is not well constrained at the redshifts of our targets, we do not expect that the true LF 
differs from our assumed LF by such gross margins, and therefore we conclude that uncertainties in our computed $B_{gh}$ statistics 
due to the assumed LF are at most $10\%$.

The second of these error sources is the method used to compare the data to the luminosity function in calculating the 
$B_{gh}$ statistic. This can be done in two ways. The data can be $k$-corrected to the rest-frame $K_{s}$-band, and 
compared to the $K_{s}$-band luminosity function given by \citet{poz}, or the data can be compared directly to the J band 
luminosity function presented by \citet{poz}, without applying a $k$-correction, as observed frame $K_{s}$-band 
approximately samples rest-frame J band at the redshifts of our sample. Both methods have advantages and disadvantages; 
the rest-frame $K_{s}$-band is a less contaminated tracer of evolved stellar mass than the J band, but suffers from the 
extra uncertainties introduced by applying $k$-corrections. Although we chose to $k$-correct our data to the rest-frame 
$K_{s}$-band, we also examined the effect of comparing our data directly to the rest-frame J-band luminosity function, 
without applying $k$-corrections. We computed $B_{gh}$ and $\sigma_{B_{gh}}$ without applying $k$-corrections, and 
using the J-band luminosity function from \citet{poz}, and found that the $B_{gh}$ values were within 1$\sigma$ of 
the values computed using $k$-corrections and the $K_{s}$-band luminosity function for all the objects except 
F10119+1429, which was within 1.5$\sigma$. Furthermore, each object still resided within the same type of environment 
as before, with P09104+4109, F10026+4949 and F10119+1429 residing in clusters (with the same significance of detection 
of clustering) and the other 3 objects lying in the field. The values of $B_{gh}$ and $\sigma_{B_{gh}}$ were however 
in all cases slightly higher using the J-band luminosity function and no $k$-corrections, with F10026+4949 and 
F10119+1429 predicted to (just) lie in Abell 2 clusters. We conclude that using the $K_{s}$-band luminosity 
function with $k$-corrections, or just the J-band luminosity function, will not significantly change our results, and 
that our choice of using the $K_{s}$-band luminosity function with $k$-corrections was the most conservative. 

\begin{figure}
\rotatebox{0}{
\centering{
\scalebox{0.41}{
\includegraphics*[18,144][592,718]{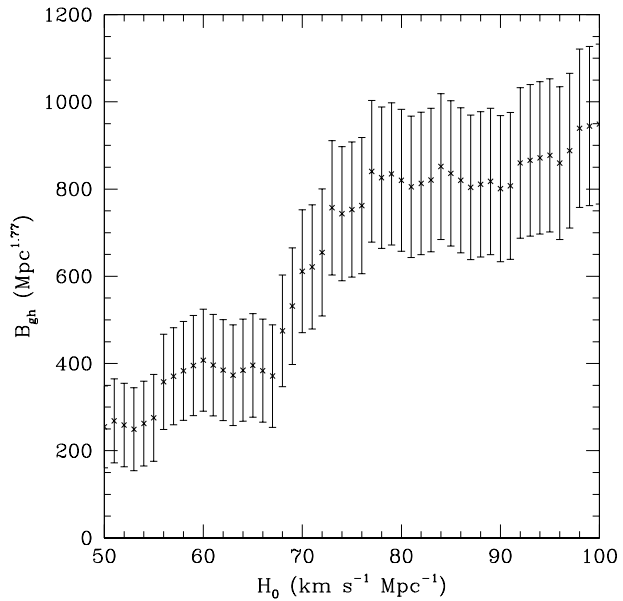}
}}}
\caption
{
The clustering amplitude of F10026+4949 as a function of $H_{0}$ (for $\Lambda = 0.7$, $\Omega_0 = 0.3$). Error bars are the Poisson 
errors.
\label{hubble_effect}
}
\end{figure}

The third of these error sources is the choice of cosmology. Whilst both $B_{gh}$ and $N_{0.5}$ are {\it relatively} insensitive 
to the choice of $\Lambda$ and $\Omega_0$, the sensitivity to $H_{0}$ is marked, particularly in the range $65<H_{0}<80$. This is 
illustrated in Figure \ref{hubble_effect}, where we have plotted $B_{gh}$ as a function of $H_{0}$ for F10026+4949. By varying $H_{0}$ 
over a relatively small range, the effect on $B_{gh}$ is dramatic, changing from $B_{gh}\sim350$ for $H_{0} = 65$ to 
$B_{gh}\sim800$ for $H_{0}\geq75$. Most current measurements of $H_{0}$ produce values in the range 70 to 75, albeit 
with a significant error (\citet{fre} and references therein). Our adopted value of $H_{0}=70$, and therefore our derived clustering 
amplitudes, are therefore conservative.

In summary, we conclude that, whilst there are further, significant sources of error than the Poisson errors on the derived clustering 
statistics, none of these sources of error should significantly change our results. Furthermore, we have in all cases adopted the most 
conservative method possible in computing the clustering amplitudes, making it highly unlikely that any of our objects that we infer 
to lie in clusters actually reside in the field.

\section{Discussion}

\subsection{The hyperluminous phenomenon at low and high redshift}
Our sample of five objects (excluding P09104), although small, exhibit a wide variety of environments, 
from poor environments to Abell $\sim1$ clusters. This strongly suggests that there is a wide variety 
of environments amongst objects with high levels of IR emission at $z\sim1$ generally, although a 
larger sample would be required to confirm this. This variety of environments is not seen amongst 
local ULIRGs, which are generally not found in rich environments \citep{san2}. The mean environmental richness of our sample, at 
$\langle B_{gh} \rangle=190\pm 45$Mpc$^{1.77}$, is higher, at just over $3\sigma$ significance, than 
the galaxy-galaxy correlation statistic 
both locally \citep{lov,guz} and at moderate redshifts \citep{hud}, though we note that the latter 
study measured values of $B_{gg}$ up to redshifts of only $z\sim0.5$. We infer that, in going from 
$z\sim0$ to $z\sim1$ the environments of IR-luminous galaxies become more diverse, and that the mean 
environment becomes richer than that of normal galaxies both locally and at moderate redshifts. 
This is supportive of the idea that a wider variety of galaxy formation processes are important 
amongst the IR-luminous galaxy population at high redshift than locally, such as hierarchical buildup or encounters 
in clusters. Indeed, one object in our sample, IRAS F10026+4949, may be an example of such a galaxy. 
This source harbours both a starburst and an AGN \citep{far3}, and lies in a rich cluster. From HST imaging, this 
source also possesses multiple very close companions \citep{far2}. It is thus an excellent candidate for being a cD 
galaxy in the process of formation in a cluster at $z=1.12$, and may be the higher redshift analogue 
of the clustered IR-luminous active galaxies P09104+4109 and P18216+6419 \citep{schn,hin,wol3}. 

\subsection{Comparison to Quasar environments}
Since environments are independent of orientation, we can compare the environments of our sample to those 
of other classes of active galaxies at comparable redshifts to examine possible relationships between HLIRGs 
and other AGN classes. The environments of AGN over a wide redshift range have been studied by many authors 
in efforts to disentangle the myriad AGN taxonomy. Radio Loud QSOs (RLQs) and Radio Galaxies (RGs) are found 
in a diverse range of environments, from the field to Abell Class 2 and greater, at both low and high redshift 
\citep{los,yee0,pre1,pre2,hil,all,zir,wol1,sng}. 
Overall, RLQs and RGs appear to prefer moderately rich environments on average, of around Abell Class 0. 
Determining whether or not there is evolution in the environments of RLQs or RGs with redshift is difficult, 
due to gravitational lensing and selection biases, but from recent results it appears that there is no significant 
difference in RLQ and RG environments at $z<1.0$ \citep{wol1}. For Radio Quiet Quasars (RQQs) a broadly 
comparable picture has now emerged. RQQs are found in a similarly diverse range of environments to RLQs, from 
the field to Abell class 2 or richer, over a wide redshift range \citep{yee0,dun0,fis,del,tan,mcl}. Indeed, the most recent results, based on 
deep optical imaging, show that the environments of RQQs and RLQs at moderate redshifts 
are statistically indistinguishable \citep{wol2}. At $z\geq1$ there is as yet no clear consensus, but the 
environments of RLQs and RQQs at these redshifts do not appear to be significantly different \citep{hint,hut0,hut1}. 

\begin{figure}
\rotatebox{0}{
\centering{
\scalebox{0.41}{
\includegraphics*[18,144][592,718]{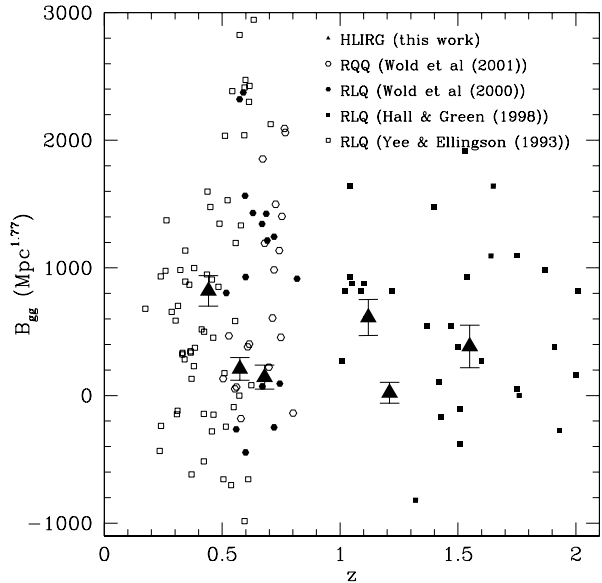}
}}}
\caption
{
The clustering amplitude vs. redshift for the HLIRGs in our sample. Also plotted are clustering amplitudes for QSOs 
taken from \citet{yee1}, \citet{hgr}, \citet{wol1} and \citet{wol2}, converted to our cosmology. The HLIRGs are plotted 
as larger symbols for clarity.
\label{qsocomp}
}
\end{figure}

In the local Universe, it was initially suggested that ULIRGs as a class are precursors to optically selected 
QSOs, although later results show that this is probably true for only a small subset of the ULIRG population 
\citep{san1,far1,tac}. This is reflected in their environments, as locally ULIRGs and QSOs reside in different 
environments, with ULIRGs generally lying in the field and QSOs lying in moderately rich environments on 
average, with a diverse range. We can compare the range of environments in our sample of HLIRGs to those of 
QSOs at comparable redshifts to see if this is also true at high redshift. The diverse range of environments 
seen in our sample of HLIRGs is qualitatively similar to the range seen in RLQs and RQQs, and our mean value 
for $B_{gh}$ is comparable to the mean galaxy-quasar correlation statistic, $B_{gq}$, for RLQs and RQQs at 
similar redshifts \citep{wol1,wol2}. This is illustrated in Figure \ref{qsocomp}, where we plot clustering 
amplitudes vs. redshifts for our HLIRGs, together with a representative sample of clustering amplitudes for QSOs 
from the literature \citep{yee1,hgr,wol1,wol2}, converted to our cosmology. The error bars on the QSOs have been 
intentionally omitted for clarity, but are comparable in size to the error bars on the HLIRGs. We note that the 
clustering measures were derived using different data, with our clustering amplitudes derived from $K_{s}$-band 
data and the clustering amplitudes from the other samples derived from multiband optical data, however it is 
unlikely that this will introduce any systematic biases between the datasets. Therefore, we extrapolate from 
this that evolutionary links between IR-luminous galaxies and optically selected QSOs may become stronger with 
increasing redshift. Furthermore, we postulate that, at approximately $z>0.5$, the range of galaxy evolution processes 
driving IR-luminous galaxy evolution caused by the greater diversity in environments is similar to the range 
driving QSO evolution. We note that one object in our sample, LBQS 1220+0939, is a QSO from the Large Bright 
Quasar Survey \citep{hfc} that was only later discovered to be hyperluminous in the IR; the selection criteria 
for this object therefore make it unsuitable for determining evolutionary connections between HLIRGs and QSOs. 
Even with LBQS 1220+0939 removed however, the sample still contains the same diverse range of environments, 
and the mean clustering amplitude is still comparable to that of QSOs at similar redshifts. 

We can take this one step further via a comparison of HLIRG and optically selected QSO host galaxies. QSOs at 
all redshifts are thought to lie in luminous, $>L^{*}$ host galaxies, with elliptical hosts becoming more prevalent 
for both RLQs and RQQs with increasing optical nuclear luminosity \citep{mcl0,per,dun1}. Conversely, although local 
ULIRGs are thought to evolve into ellipticals with a few also passing though a QSO phase, are 
thought to be mergers between (optically) sub-$L^{*}$ galaxies, with the majority of ULIRGs themselves also being 
sub-$L^{*}$ \citep{col}. For the HLIRGs in our sample, this is not the case. One object, F14218+3845, is a QSO with 
a host magnitude that suggests (similarly to the other QSOs in the sample) the host is a very luminous, massive 
system \citep{far2}. For the two non-QSOs in our sample, the picture is less clear, with one object (F00235+1024) 
being 0.5 magnitudes dimmer than $m^{*}_{K}$, but the other (F10026+4949) over a magnitude brighter than $m^{*}_{K}$. 
We note however that the $I$-band magnitudes of those HLIRGs without a QSO presented by \citet{far2} are, with 
the exception of F00235+1024, much brighter than $m_{I}^{*}$. Whilst more data is evidently needed, it appears 
reasonable to assume that IR-luminous galaxies may become more similar to QSO hosts with increasing redshift. 
Whilst this reinforces the idea that evolutionary links between IR-luminous galaxies and QSOs become stronger 
with increasing lookback time, it also suggests that, at $z\geq0.5$, a greater number of IR-luminous galaxies 
evolve directly into optically selected QSOs than locally.

\section{Conclusions}
We have presented deep wide-field K-band imaging of the fields of six Hyperluminous Infared Galaxies, and quantified 
their environments using the $B_{gh}$ galaxy-HLIRG correlation amplitude, and the $N_{0.5}$ clustering statistic. We 
conclude the following:

\noindent 1) The HLIRGs in our sample reside in a diverse range of environments, from the field to Abell 2 clusters. 
The mean clustering level of the sample, at $\langle B_{gh} \rangle =190 \pm 45$, and the range of environments, are both 
significantly greater than those of the most luminous IR galaxies locally. We infer that, at high redshift, the galaxy 
evolution processes driving the evolution of IR-luminous galaxies are more diverse than at low redshift, and include 
mergers between gas-rich spirals in the field, but also include encounters in clusters and hierarchical buildup. 

\noindent 2) The mean clustering amplitude of the sample, and the range in environments, are comparable to those of 
QSOs over a similar redshift range. We postulate from this that, at $z\geq0.5$, the range of galaxy evolution processes driving 
IR-luminous galaxy evolution is similar to the range that drives QSO evolution. When combined with the similarities 
between HLIRG host galaxies and QSO host galaxies at the same redshifts, this further suggests that a greater fraction of 
IR-luminous galaxies evolve directly into optically selected QSOs at high redshift than do locally. 
 
\section*{Acknowledgments}
We thank Carol Lonsdale and Margrethe Wold for illuminating discussion, and the referee for a very helpful report. 
This paper is based on observations made with the William Herschel Telescope operated on the island of La Palma by 
the Isaac Newton Group in the Spanish Observatorio del Roque de los Muchachos of the Instituto de Astrofisica de 
Canarias. This research has made use of the NASA/IPAC extragalactic database (NED) which is operated by the Jet 
Propulsion Laboratory, California Institute of Technology, under contract with the National Aeronautics and Space 
Administration.DF was supported by NASA grant NAG 5-3370 and by the Jet Propulsion Laboratory, California Institute 
of Technology, under contract with NASA. MF and AK were supported by PPARC.

\end{document}